\documentclass{webofc}

\usepackage[varg]{txfonts}   
\hypersetup{colorlinks=true,citecolor=blue,urlcolor=blue,linkcolor=blue}
\usepackage{tikz}
\usetikzlibrary{shapes,arrows,positioning,fit,backgrounds,calc}
\usepackage{caption}
\usepackage{subcaption}
\usepackage{float}

\begin{document}
\title{Double Deeply Virtual Compton Scattering as a probe of Generalized Parton Distributions}
\author{\firstname{Juan Sebastian} \lastname{Alvarado}\inst{1}\fnsep\thanks{\email{jsalvaradog@outlook.com}} \and
        \firstname{Mostafa} \lastname{Hoballah}\inst{1}\fnsep\and
        \firstname{Eric} \lastname{Voutier}\inst{1}\fnsep
}
\institute{Université Paris-Saclay, CNRS/IN2P3, IJCLab, 91405 Orsay, France}
\abstract{Generalized Parton Distributions (GPDs) are multidimensional structure functions of hadrons, encoding mechanical and spin properties through the correlation of the momentum and transverse position of partons. While channels like Deeply Virtual Compton Scattering (DVCS) access GPDs in a constrained kinematic domain, Double Deeply Virtual Compton Scattering (DDVCS) offers broader access to the GPD phase space, though it remains unmeasured. We present a feasibility study of DDVCS observables using polarized electron beams at Jefferson Lab and the EIC, focusing on their sensitivity to helicity-conserving GPDs and implications for Compton Form Factor (CFF) extraction. Moreover, we provide experimental projections supporting the feasibility of the measurements at both facilities.
}

\maketitle
\section{Introduction}
\begin{figure}[H]
    \centering
    \begin{subfigure}[b]{0.3\textwidth}
        \centering
        \includegraphics[width=0.7\textwidth]{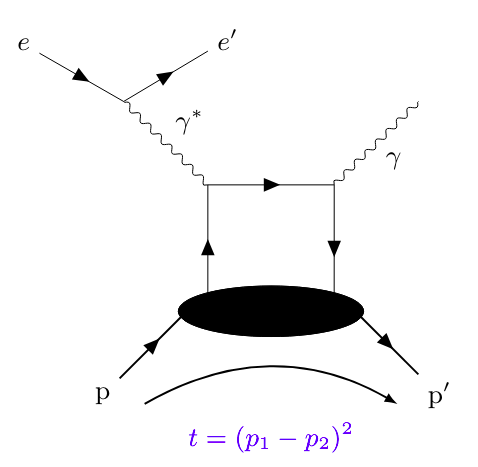}
    \end{subfigure}
    \begin{subfigure}[b]{0.3\textwidth}
        \centering
        \includegraphics[width=0.7\textwidth]{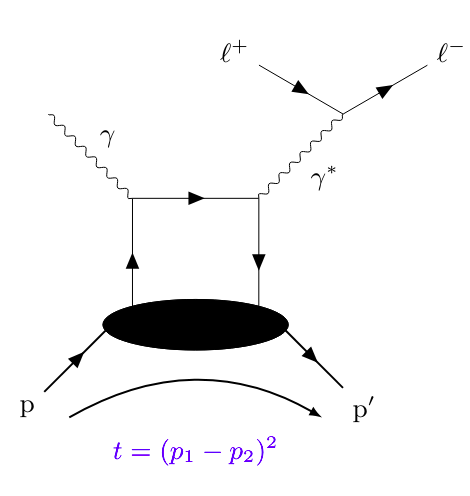}
    \end{subfigure}
    \begin{subfigure}[b]{0.3\textwidth}
        \centering
        \includegraphics[width=0.9\textwidth]{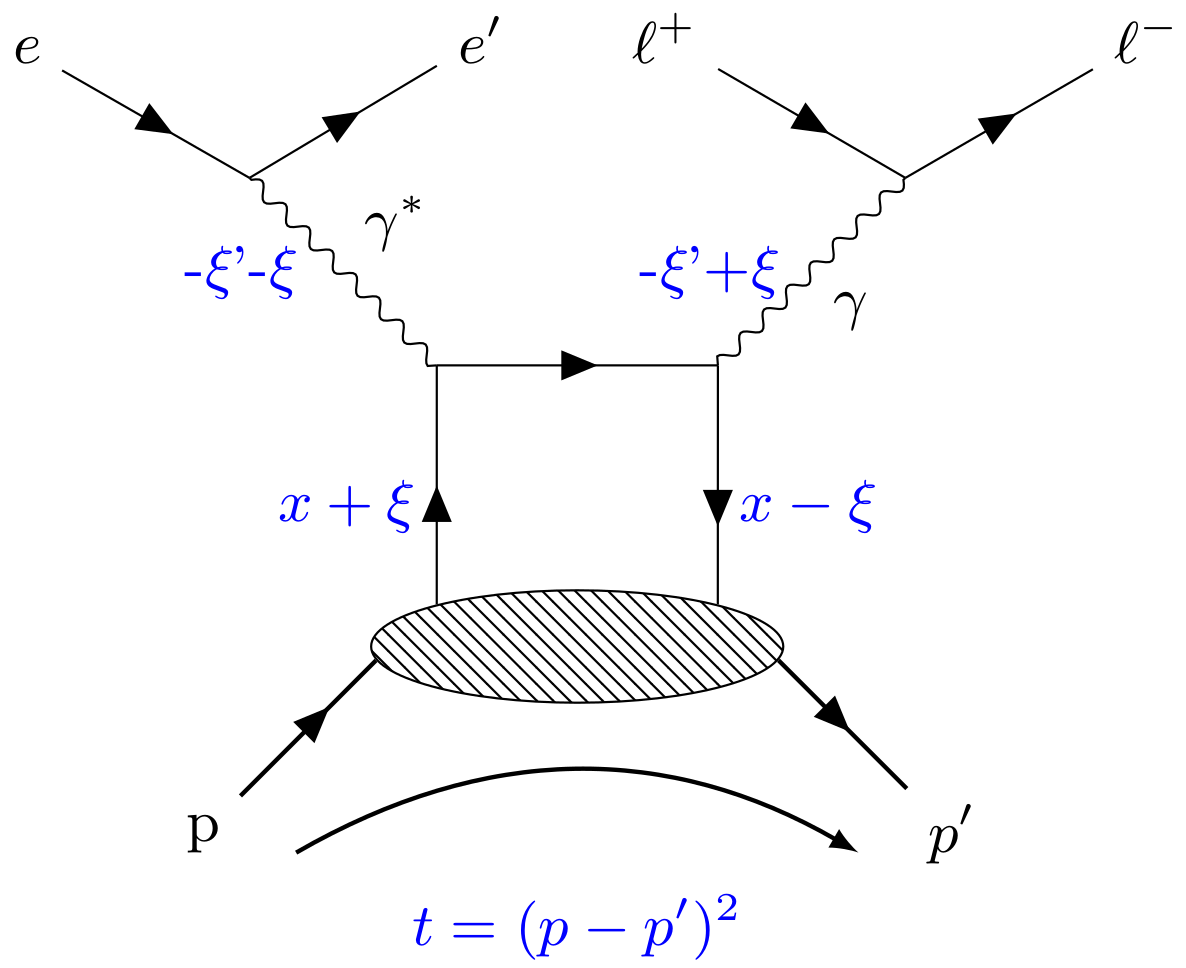}
    \end{subfigure}
    \caption{DVCS (left), TCS (middle) and DDVCS (right) reactions.}
    \label{DVCS_TCS}
\end{figure}

Generalized Parton Distributions (GPDs) provide a unified description of the nucleon’s internal structure, encoding nucleon properties arising from the correlation between the longitudinal momentum and transverse spatial position of partons \cite{GPD4}. GPDs are probed in a restricted kinematic domain through deep exclusive reactions like Deeply Virtual Compton Scattering (DVCS) and Time-like Compton Scattering (TCS) (Fig. \ref{DVCS_TCS}). However, Double DVCS (DDVCS)\cite{DDVCS1, DDVCS2} stands as a promising channel granting access to regions in the GPD phase space otherwise inaccessible. This is possible through experimental observables that are directly interpretable in terms of Compton Form Factors (CFFs), which involve the four chiral-even GPDs of a spin $1/2$ particle.

\section{Experimental configuration}
Due to its lower cross section and di-lepton final state, DDVCS measurements require a high luminosity and a large acceptance detector with muon detection capabilities. At Jefferson Lab (JLab), the proposed $\mu$CLAS12 and SoLID$\mu$ spectrometers aim to provide access to the valence region of GPDs through Beam Spin Asymmetry (BSA) and muon-Charge Asymmetry ($\mu$CA) measurements, accessing the imaginary and real parts of CFFs respectively, leveraging their projected muon detection capabilities and support for luminosities of to $10^{37}$ cm$^{-2}$s$^{-1}$. In parallel, access to the sea region of GPDs is foreseen at the future Electron-Ion Collider (EIC), provided a DDVCS identification from potential muon detection capabilities.
\section{Sample of projected measurements}
\begin{figure}[H]
    \centering
    \begin{subfigure}[b]{0.23\textwidth}
        \centering
        \includegraphics[width=\textwidth]{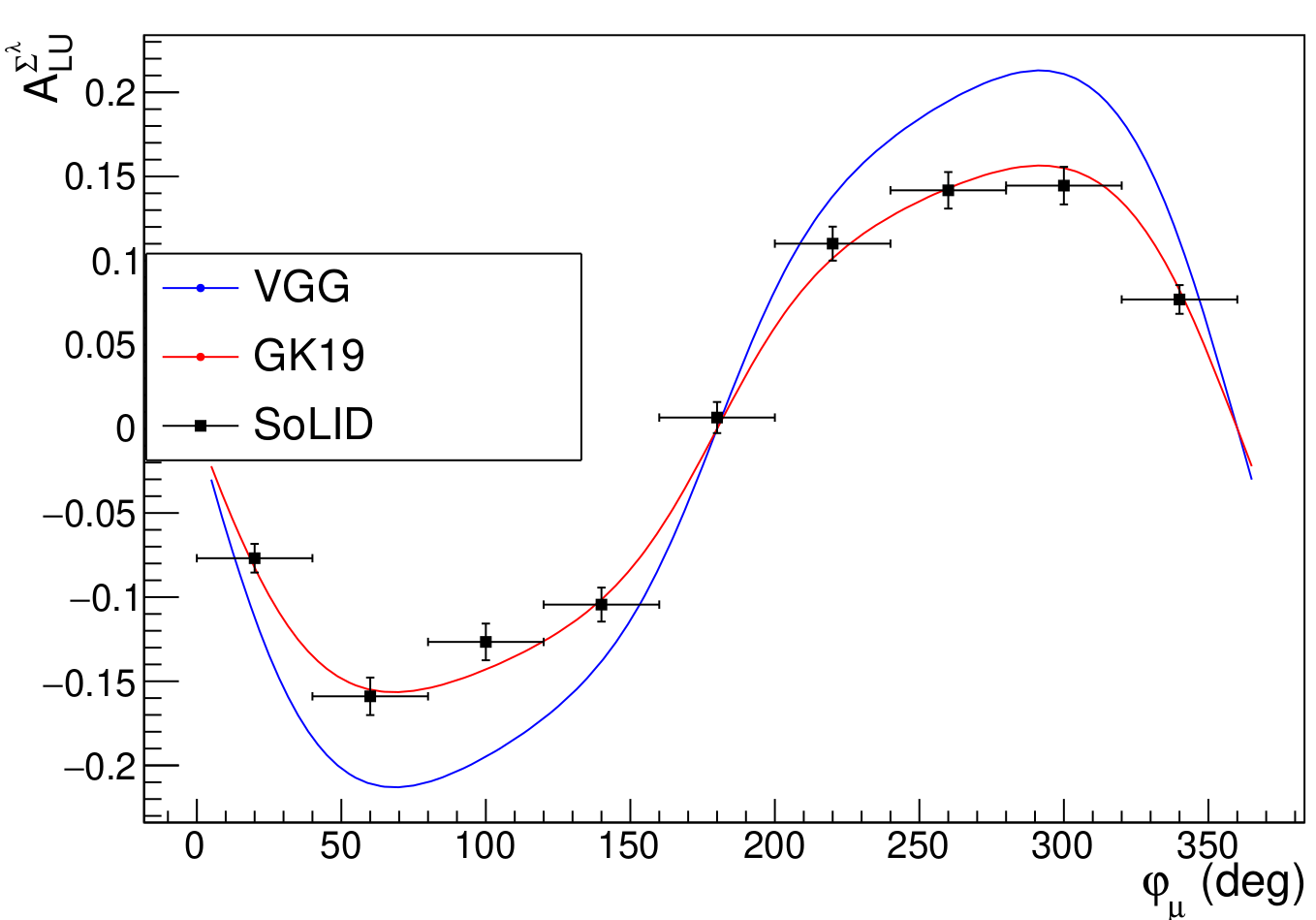}
        \caption{BSA at JLab.}
    \end{subfigure}
    \begin{subfigure}[b]{0.23\textwidth}
        \centering
        \includegraphics[width=\textwidth]{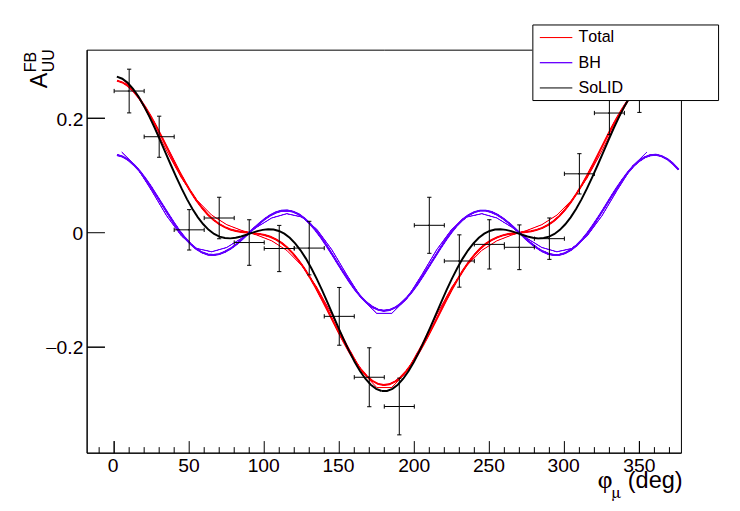}
        \caption{$\mu$CA at JLab.}
    \end{subfigure}
    \begin{subfigure}[b]{0.23\textwidth}
        \centering
        \includegraphics[width=\textwidth]{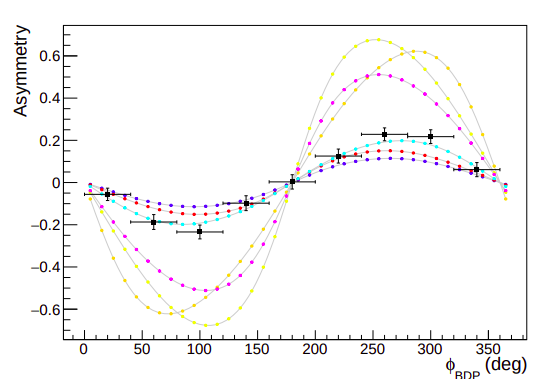}
        \caption{BSA at EIC.}
    \end{subfigure}
    \begin{subfigure}[b]{0.23\textwidth}
        \centering
        \includegraphics[width=\textwidth]{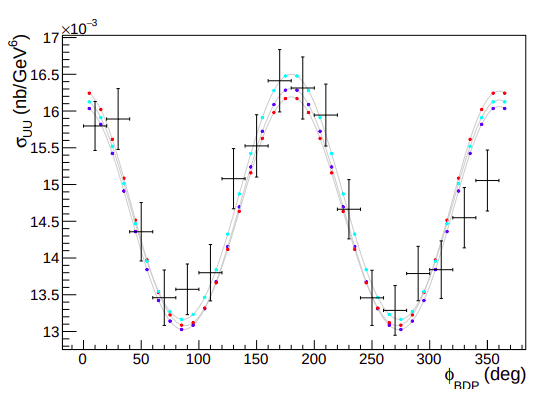}
        \caption{$\sigma_{UU}$ at EIC.}
    \end{subfigure}
    \caption{Sample experimental projections.}
    \label{Projections}
\end{figure}

Experimental projections support the feasibility of the DDVCS measurements at JLab and the EIC  (see \cite{Myself} for details). Fig. \ref{Projections} shows an example of the foreseen measurements within the SoLID$\mu$\cite{Myself, SoLIDu} and EIC experimental configurations. At JLab, precise BSA and $\mu$CA measurements are foreseen given the large amplitudes expected from model predictions, allowing for discrimination among models. In contrast, model predictions at EIC kinematics suggest smaller, yet measurable, BSA amplitudes. Additionally, cross section measurements can also be foreseen at EIC, further expanding its sensitivity to GPD-related observables. 

\noindent
\textbf{Acknowledgements}: This work was supported by the Programme blanc of the physics Graduate School of the PHENIICS doctoral school [Contract No. D22-ET11] and the french Centre National de la Recherche Scientifique (CNRS).

\end{document}